\documentclass[twoside]{LCWS11}
\usepackage[latin1]{inputenc}
\usepackage[dvips]{graphicx,epsfig,color}
\usepackage{wrapfig,rotating}
\usepackage{amssymb,amsmath,array}
\usepackage{cite}
\def\lromn#1{\uppercase\expandafter{\romannumeral#1}}

\def\Slash#1{{\ooalign{\hfil$#1$\hfil\crcr\hfil$/$\hfil}}}

\begin{document}


\title{
Higgs portal dark matter at a linear collider }
\author{Takehiro Nabeshima$^1$
\vspace{.3cm}\\
Department of Physics, University of Toyama, Toyama 930-8555, Japan\\}

\maketitle

  \abstract{
\hspace{12pt}We investigate the possibility of detecting dark matter 
at TeV scale linear colliders in the scenario where 
the dark matter interacts with standard model particles 
only via the Higgs boson.
In this scenario, the dark matter would be difficult to be tested 
at the LHC especially when the decay of the Higgs boson into 
a dark matter pair is not kinematically allowed.
In this talk, we discuss whether even such a case can be
explored or not at the ILC and CLIC via the Z boson fusion process. 
This talk is mainly based on 
Refs.~\cite{Kanemura:2010sh} and \cite{Kanemura:2011nm}.
} 

\section{Introduction}

\hspace{12pt}The existence of dark matter has been established. It occupies 
more than one fifth of the energy density in the Universe~\cite{Komatsu:2010fb}.
If the essence of dark matter is a particle, 
it must be electrically neutral and weakly interacting. 
A plausible candidate for dark matter would be 
a Weakly Interacting Massive Particle (WIMP). 
According to the WMAP experiment~\cite{Komatsu:2010fb}, 
the WIMP dark matter mass is deduced at the TeV scale or below 
by a rough estimation. 
Therefore, we may be able to directly produce the dark matter 
and to test it at high energy collider experiments 
such as the CERN Large Hadron Collider (LHC), 
the International Linear Collider (ILC)~\cite{Flacher:2008zq} 
and the Compact Linear Collider (CLIC)~\cite{Accomando:2004sz}.

\hspace{12pt}The fact that the scale of the WIMP dark matter mass is similar to 
the electroweak symmetry breaking scale would indicate 
that there is a strong connection between the Higgs boson and the dark matter.
There are many new physics models involving a dark matter candidate.
In some of them, it can happen that the dark matter couples only to the Higgs boson 
in the low energy effective theory below TeV scale, 
where the stability of dark matter is guaranteed by an unbroken discrete symmetry.
Such a scenario is often called the Higgs portal dark matter scenario
~\cite{Kanemura:2010sh, Kanemura:2011nm, higgsportal-scalar1, higgsportal-scalar3, higgsportal-scalar4, Higgsportal, higgsportal-fermion1, higgsportal-vector1}.

\hspace{12pt}In this talk
~\cite{Kanemura:2010sh, Kanemura:2011nm}, we discuss the possibility 
whether the Higgs portal dark matter can be tested at TeV scale colliders 
such as the LHC, the international Linear Collider (ILC)~\cite{Flacher:2008zq} 
and the Compact Linear Collider (CLIC)~\cite{Accomando:2004sz}.
We here study pair production processes of the Higgs portal dark matter 
via weak boson fusion processes. 
As a result, when the Higgs boson $h$ decay into pair of the dark matter $D$, $h\to DD$, is kinematically allowed, 
so that the signal would be detectable at the LHC after appropriate kinematic cuts~\cite{Eboli:2000ze} 
unless the coupling constant between $h$ and $D$ is too small. 
On the other hand, if the decay $h\to DD$ is not kinematically allowed, 
the detection of the signal would be hopeless at the LHC. 
However, 
the signal could be detected at the linear collider with
a collision energy $\sqrt{s}>1$~TeV the integrated luminosity 1 ab$^{-1}$, 
when the mass of $D$ is up to about 100 GeV and the Higgs boson mass is 120 GeV.

\section{The model}

\hspace{12pt}We here consider the simple model in which a dark matter field is added to the SM. We impose an unbroken $Z_2$ parity, under which the dark matter is assigned to be odd while the SM particles are to be even. Stability of the dark matter is guaranteed by the $Z_2$ parity. We consider three possibilities for the spin of the dark matter; i.e., the real scalar $\phi$, the Majorana fermion $\chi$ and the real massive vector $V_{\mu}$. 

\hspace{12pt}The Lagrangian for each case of 
the dark matter is given~\cite{Kanemura:2010sh} by 
\begin{eqnarray}
 {\cal L}_{\rm S}
 &=&
 {\cal L}_{\rm SM}
 +
 \frac{1}{2} \left(\partial \phi\right)^2
 -
 \frac{1}{2} M_{\rm S}^2 \phi^2
 -
 \frac{c_{\rm S}}{2}|H|^2 \phi^2
 -
 \frac{d_{\rm S}}{4!} \phi^4, 
 \label{eq:S} \\
 {\cal L}_{\rm F}
 &=&
 {\cal L}_{\rm SM}
 +
 \frac{1}{2}\bar\chi\left(i\Slash{\partial} - M_{\rm F}\right)\chi
 -
 \frac{c_{\rm F}}{2\Lambda} |H|^2 \bar\chi \chi, 
 \label{eq:F} \\
 {\cal L}_{\rm V}
 &=&
 {\cal L}_{\rm SM}
 -
 \frac{1}{4} V^{\mu\nu} V_{\mu \nu}
 +
 \frac{1}{2} M_{\rm V}^2 V_\mu V^\mu
 +
 \frac{c_{\rm V}}{2} |H|^2 V_\mu V^\mu
 -
 \frac{d_{\rm V}}{4!} (V_\mu V^\mu)^2, 
 \label{eq:V}
\end{eqnarray}
where $M_i$($i = $ S, F and V) are the bare  masses of $\phi$, $\chi$ and $V_{\mu}$, $c_i$ and $d_i$ are dimensionless coupling constants, $\Lambda$ is a dimensionfull parameter, and $V_{\mu\nu}$ and $B_{\mu\nu}$ are Abelian field strength tensors. In this case, the dark matter in Eqs.~(\ref{eq:S})-(\ref{eq:V}) only couples to the SM particles via the Higgs doublet field $H$: i.e., it is so-called the Higgs portal dark matter. 

\section{Dark Matter signals at the LHC}
\subsection{The case $m_D^{} < m_h^{}/2$}

\hspace{12pt}In this case, when the mass of the Higgs boson is not heavy ($m_h$ \hspace{0.3em}\raisebox{0.4ex}{$<$}\hspace{-0.75em}\raisebox{-.7ex}{$\sim$}\hspace{0.3em}  150~GeV), 
its partial decay width into quarks and leptons is suppressed 
due to small Yukawa couplings. 
As a result, the branching ratio of the decay into the dark matter particles 
can be almost 100\% unless the interaction between the dark matter 
and the Higgs boson is too weak~\cite{Kanemura:2010sh}. 

\hspace{12pt}There are several studies on the invisible decay of the Higgs boson at 
the LHC~\cite{higgsportal-scalar3,Eboli:2000ze}. The most significant process for investigating 
such a Higgs boson is found to be its production through 
weak gauge boson fusion. 
According to the analysis in Ref.~\cite{InvH}, 
the 30 fb$^{-1}$ data can allow us to identify the production 
of the invisibly decaying Higgs boson at the 95\% confidence level 
when its invisible branching ratio is larger than 0.250 for
$m_h = 120$ GeV. 
With the use of the analysis, we plot the experimental sensitivity 
to detect the signal in Fig.~\ref{fig:results}. The sensitivity 
is shown as green lines with $m_D^{} \leq m_h^{}/2$, where 
the signal can be observed in the regions above these lines. 
Most of parameter regions with $m_D^{} \leq m_h^{}/2$ 
can be covered by investigating the signal of the invisible decay 
at the LHC. 

\subsection{The case $m_D^{} \geq m_h^{}/2$} 

\hspace{12pt}In this case, the dark matter cannot be produced 
from the decay of the Higgs boson. We consider, however, 
the process of weak gauge boson fusion again~\cite{Kanemura:2010sh}. 
With $V^*$ and $h^*$ being a virtual weak gauge boson and the Higgs boson, 
the signal is from the process 
$qq \rightarrow qqV^*V^* \rightarrow qqh^* \rightarrow qqDD$, 
which is characterized by two energetic quark jets 
with large missing energy and a large pseudo-rapidity gap 
between them~\cite{Eboli:2000ze}. 

\hspace{12pt}Following the Ref.~\cite{Eboli:2000ze}, we apply kinematical 
cuts in order to reduce backgrounds,
\begin{eqnarray}
 &&
 p^j_T > 40~{\rm GeV},
 \qquad
 \Slash{p}_T > 100~{\rm GeV},
 \nonumber \\
 &&
 |\eta_j| < 5.0,
 \qquad
 |\eta_{j_1} - \eta_{j_2}| > 4.4,
 \qquad
 \eta_{j_1} \cdot \eta_{j_2} < 0,
 \nonumber \\
 &&
 M_{j_1j_2} > 1200~{\rm GeV},
 \qquad
 \phi_{j_1j_2} < 1,
 \label{kinematical cuts}
\end{eqnarray}
where $p^j_T$, $\Slash{p}_T$, and $\eta_j$ are the transverse momentum of $j$, 
the missing energy, and the pseudo-rapidity of $j$, respectively. 
The invariant mass of the two jets is denoted by $M_{jj}$, 
while $\phi_{jj}$ is the azimuthal angle between two jets. 
We also impose a veto of central jet activities with $p_T > 20$ GeV 
in the same manner of this reference. From the analysis of these
backgrounds, it turns out that, at the LHC with the energy of
$\sqrt{s}=14$ TeV and the integrated luminosity of 100 fb$^{-1}$,
the signal will be detected at 95\% confidence level 
when its cross section exceeds 4.8 fb after applying these kinematical cuts~\cite{Eboli:2000ze}.

\begin{figure}[t]
 \begin{center}
  \includegraphics[scale=0.15]{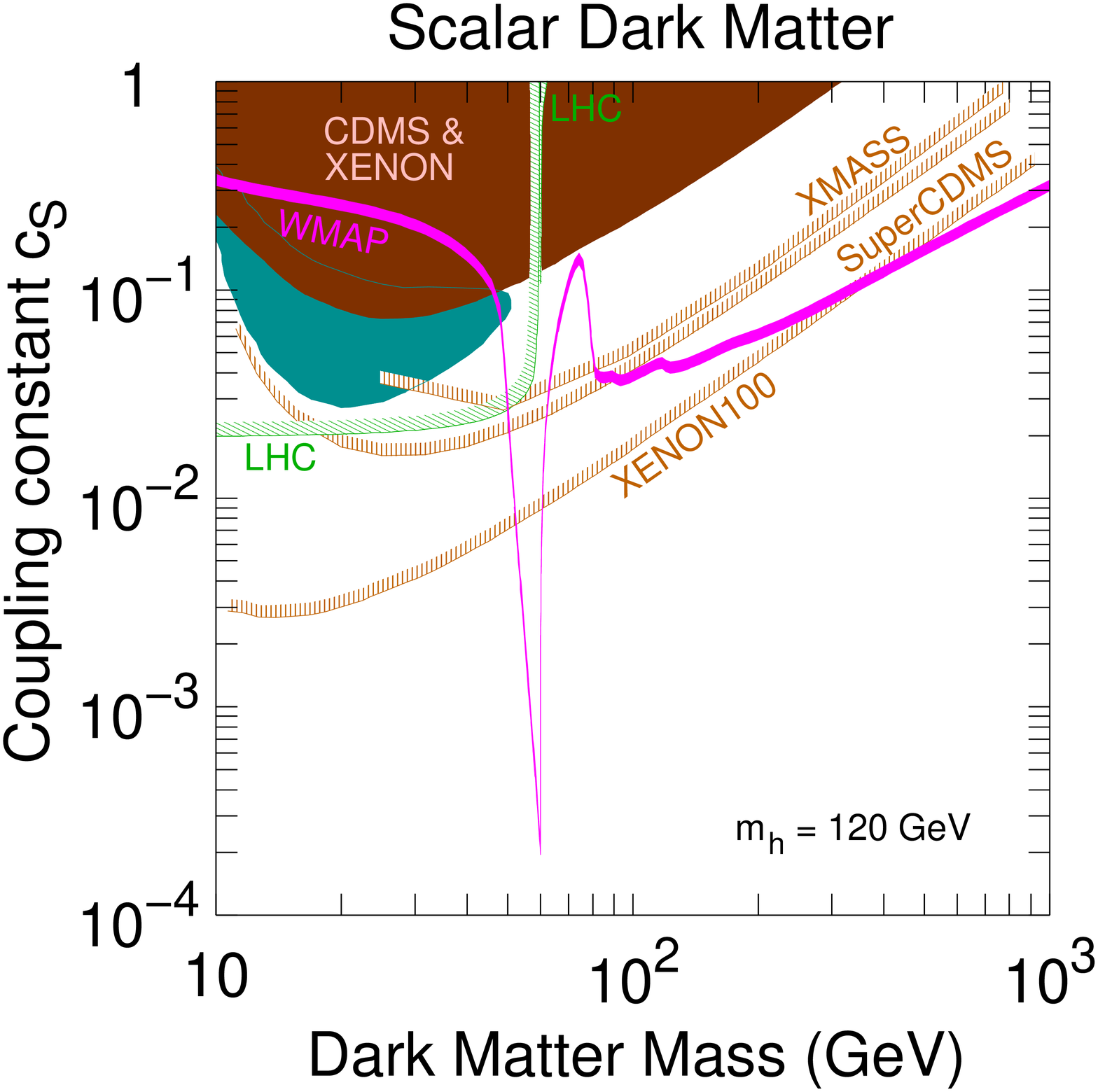}
  \includegraphics[scale=0.15]{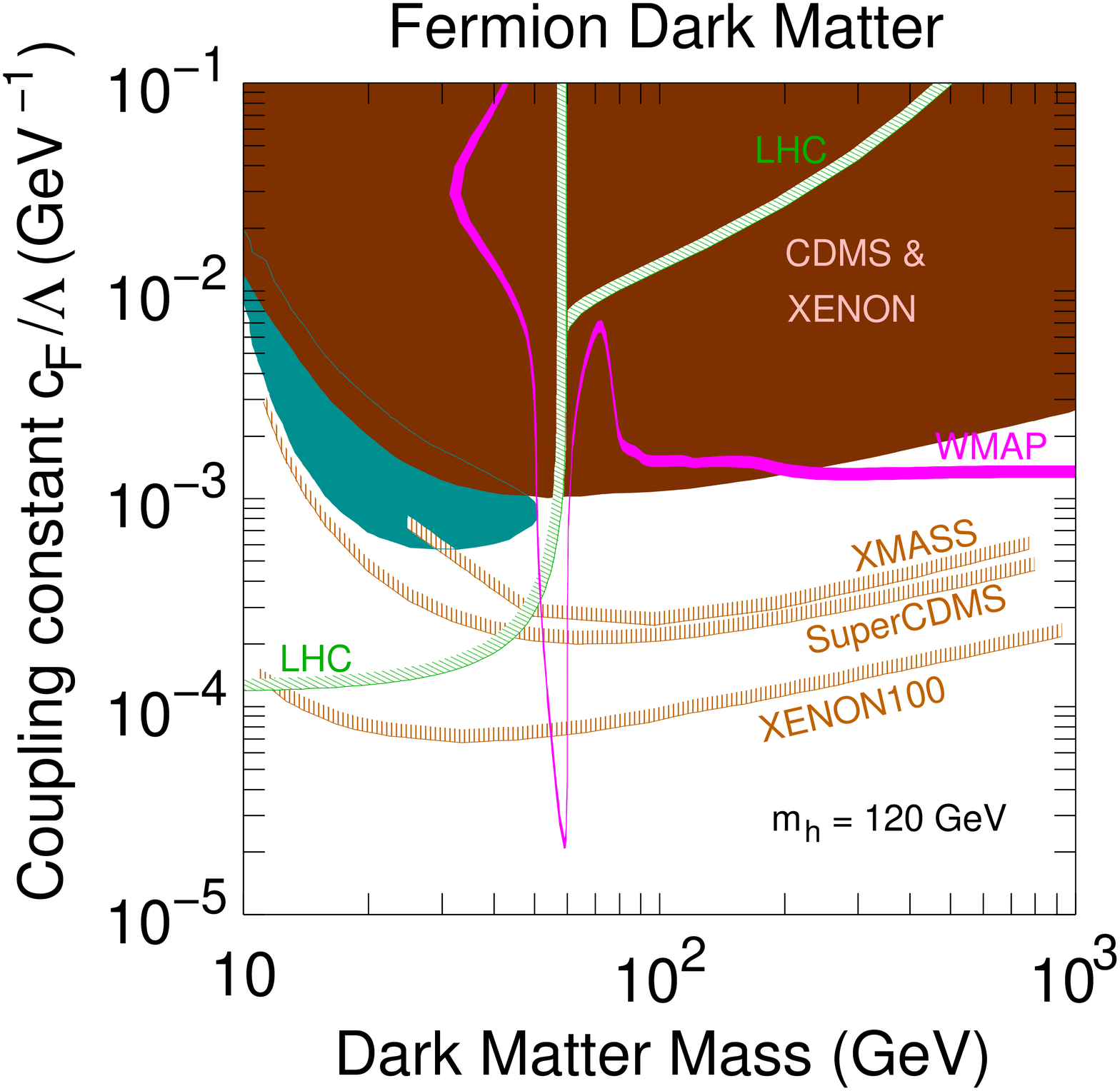}
  \includegraphics[scale=0.15]{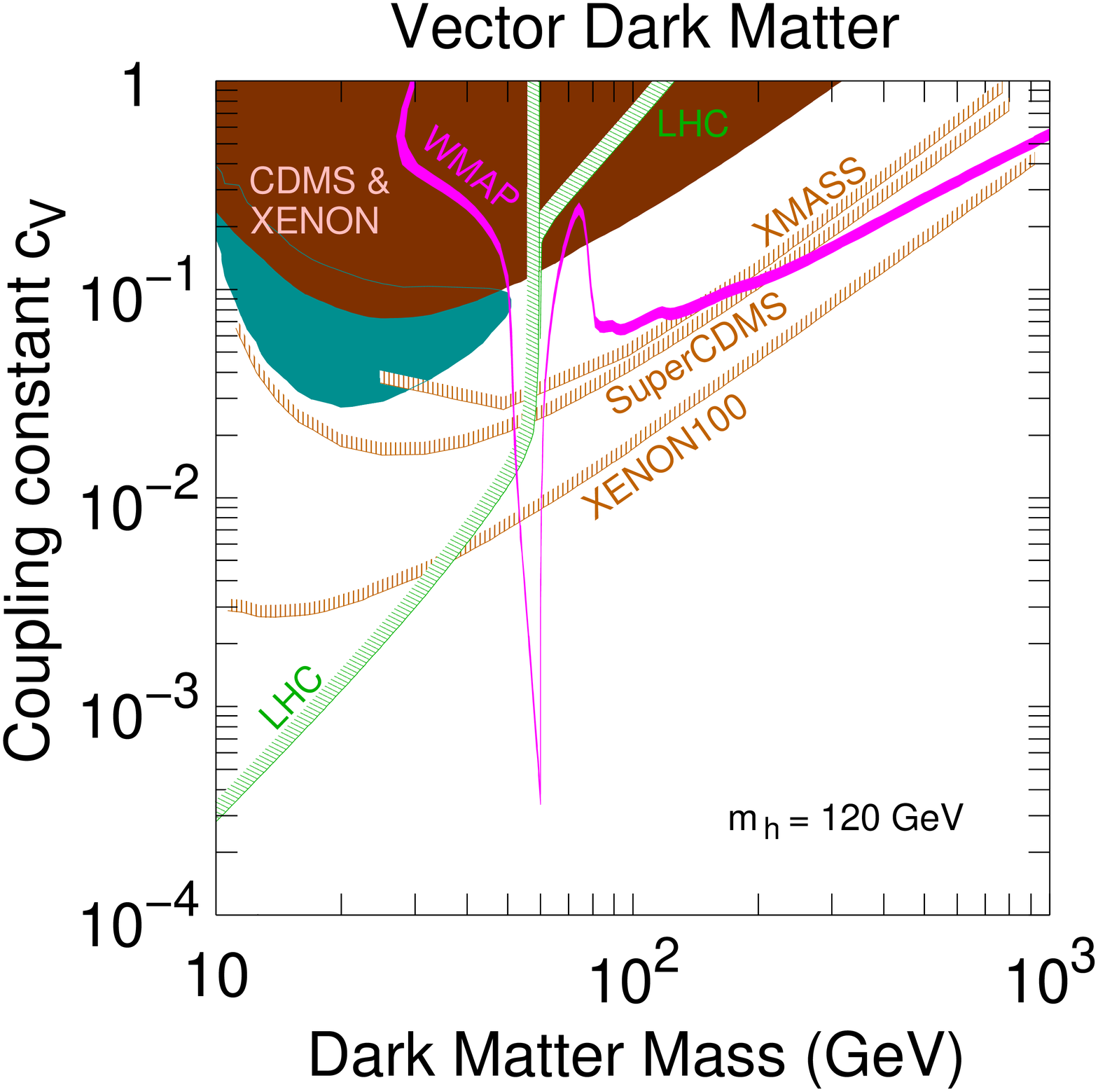}
 \end{center}
 \caption{\small 
 Sensitivities to detect the dark matter signal at the LHC. Constraints and
expected sensitivities on direct detection experiments for dark matter are also shown.}
 \label{fig:results}
\end{figure}

\hspace{12pt}The result is shown 
in Fig.~\ref{fig:results} as the regions above green lines for $m_D^{} \geq m_h^{}/2$, 
where with an integrated luminosity of 100 fb$^{-1}$ 
the signal at 95\% confidence level can be observed.
The allowed region which satisfies the WMAP data (2$\sigma$) is the magenta region. 
The sensitivity does not reach the region 
consistent with the WMAP observation, but it is close 
for fermion and vector dark matters with $m_h = 120$ GeV. 

\section{Dark Matter signals at the $e^+ e^-$collider}

\hspace{12pt}We consider the possibility to detect the dark matter at electron-positron linear colliders such as the ILC and the CLIC~\cite{Kanemura:2011nm}. We are interested in the case of $m_h < 2m_D$, where the Higgs boson cannot decay into a pair of dark matters. We concentrate on the $Z$ boson fusion process $e^+e^- \to e^+e^-Z^*Z^* \to e^+e^-h^* \to e^+e^-DD$. This process can, in principle, be used to detect the dark matter by measuring the outgoing electron and positron in the final state and by using the energy momentum conservation.

\hspace{12pt}We impose the polarization for both incident electron and positron beams~\cite{Flacher:2008zq}; 
\begin{eqnarray}
\frac{N_{e^-_R}-N_{e^-_L}}{N_{e^-_R}+N_{e^-_L}} = 80\%,
\hspace{1cm}
\frac{N_{e^+_R}-N_{e^+_L}}{N_{e^+_R}+N_{e^+_L}} = 50\%,
\end{eqnarray}
where $N_{e^-_{R,L}}$ and $N_{e^+_{R,L}}$ are numbers of right (left) handed electron and positron in the beam flux per unit time. By using the polarized beams, the backgrounds which are mediated by the $W$ boson can be reduced. The backgrounds mediated by the $Z$ boson are reduced by the basic cut in Eq.~(\ref{eq:Minv}) as we will see soon.

\hspace{12pt}The cross section of the signal process is the larger as the collision energy $\sqrt{s}$ increases, 
so that the higher collision energy may be more useful to detect the signal. 
However, for $\sqrt{s}= 1$-$5$~TeV, the outgoing electron and positron tend to be emitted to forward and backward 
directions, and the detectability of the leptons near the beam line is therefore essentially important. 
We assume the detectable area as~\cite{Bambade:2006qc}
\begin{eqnarray}
 |\cos\theta|
 <
 0.9999416, 
 \label{eq:cos}
\end{eqnarray}
where $\theta$ is the scattering angle. Assuming the situation that the Higgs boson mass is already known, we impose the condition for the missing invariant mass $M_{\rm inv}$ as 
\begin{eqnarray}
 M_{\rm inv}
 >
 m_h^{}, 
 \label{eq:Minv}
\end{eqnarray}
in order to discuss the detection of the dark matter in the case $m_D^{} > m_h^{}/2$. 

\hspace{12pt}Backgrounds against the signal process are all the process with the final state of $e^+e^-$ with a missing momentum. The main background processes are those with the final state $e^+e^-\nu_e\overline{\nu}_e$. 
We impose kinematical cuts in order to gain the signal significance~\cite{Kanemura:2011nm}. 
\begin{eqnarray}
 E_{\rm inv}
 <
 0.4\sqrt{s}~{\rm GeV},
 \ \ \ \  \phi_{ee}
 <
 2.3~{\rm rad},\ \ \ \ \ \ \ \ \ \ \ \ \ \ \ \ \ \ 
 \nonumber \\
 M_{ee}^{S1}
 >
 600~{\rm GeV},
 \ \ \ \ 
 M_{ee}^{F1}
 >
 600~{\rm GeV},
 \ \ \ \ 
 M_{ee}^{V1}
 >
 600~{\rm GeV},
 \nonumber \\
 M_{ee}^{S5}
 >
 4200~{\rm GeV},
 \ \ \ \ 
 M_{ee}^{F5}
 >
 3900~{\rm GeV},
 \ \ \ \ 
 M_{ee}^{V5}
 >
 3000~{\rm GeV},
 \label{Eq:Mee}
\end{eqnarray}
where $E_{\rm inv}$, $M_{ee}$ and $\phi_{ee}$ are the missing energy, 
the invariant mass depending on the spin of the dark matter for $\sqrt{s} = 1$~TeV and 5~TeV and 
the azimuthal angle between outgoing electron and positron, respectively.
As a result, the significance to detect the signal,  
${N_S}/{\sqrt{N_S+N_B}}$ with $N_{S}$ ($N_{B}$) being the event number for signal (backgrounds), 
can be greater than one even if the Higgs boson cannot decay into a dark matter pair.
\begin{figure}[t]
 \begin{center}
  \includegraphics[scale=0.25]{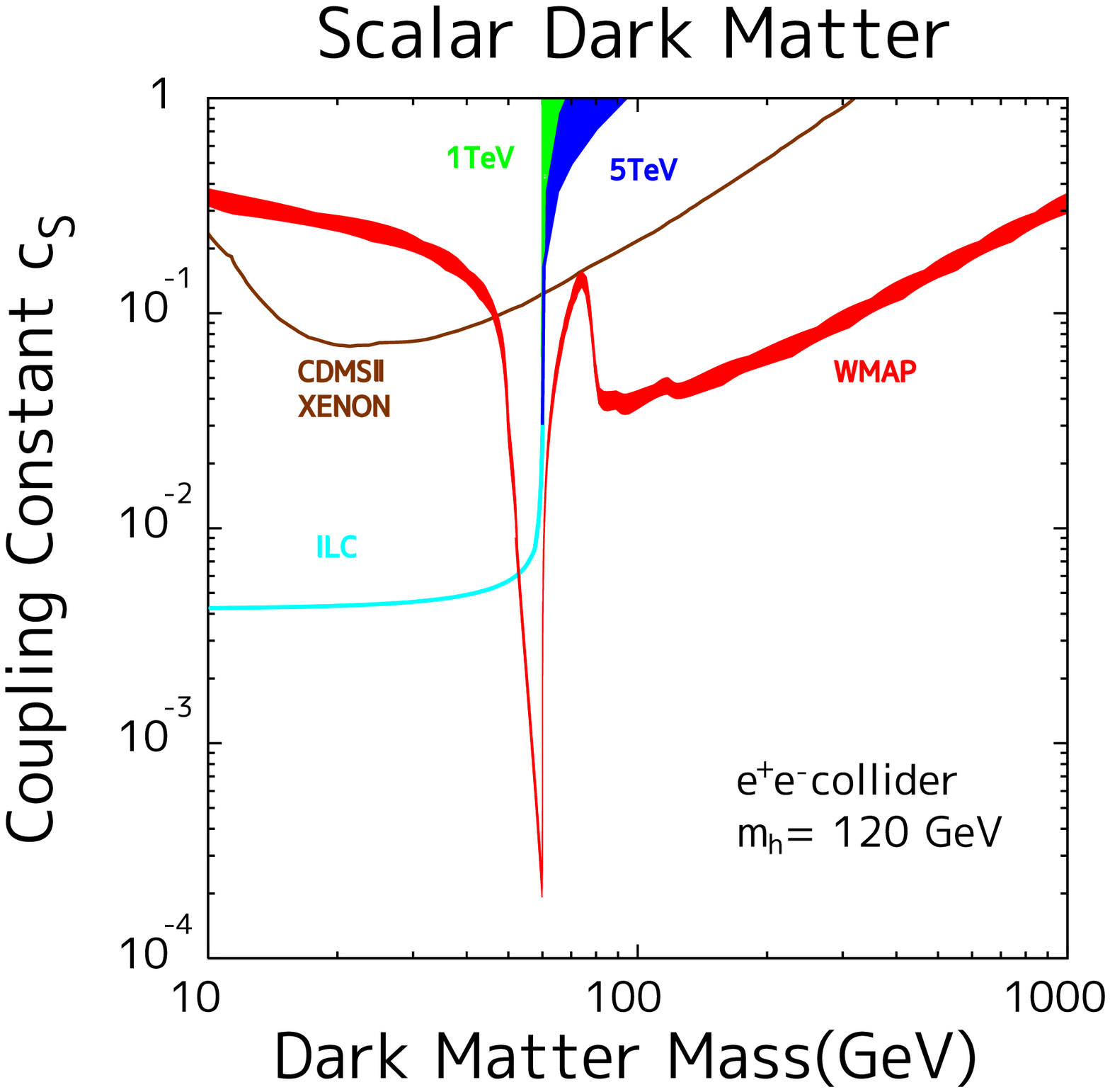}
  \includegraphics[scale=0.25]{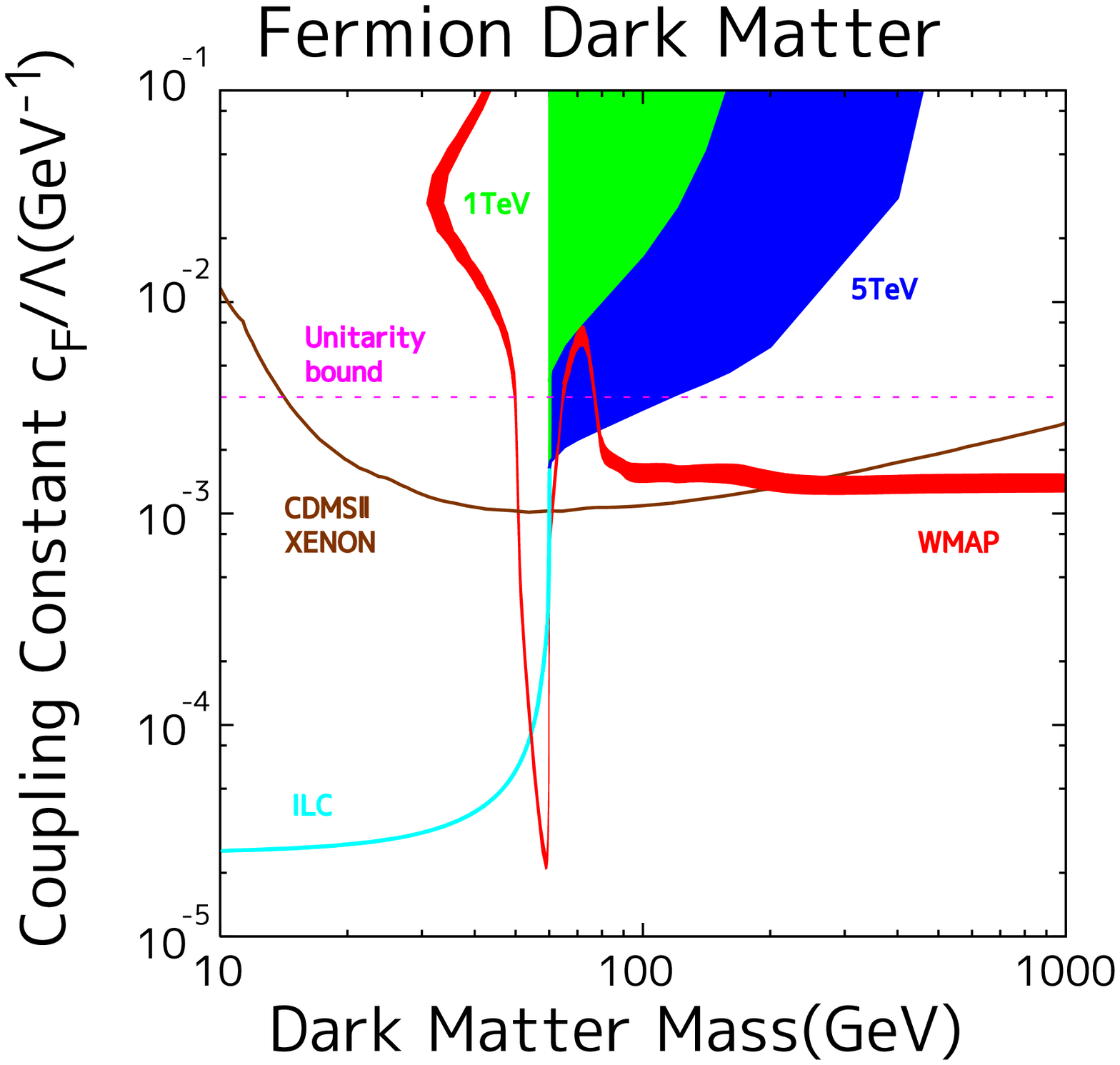}
  \includegraphics[scale=0.25]{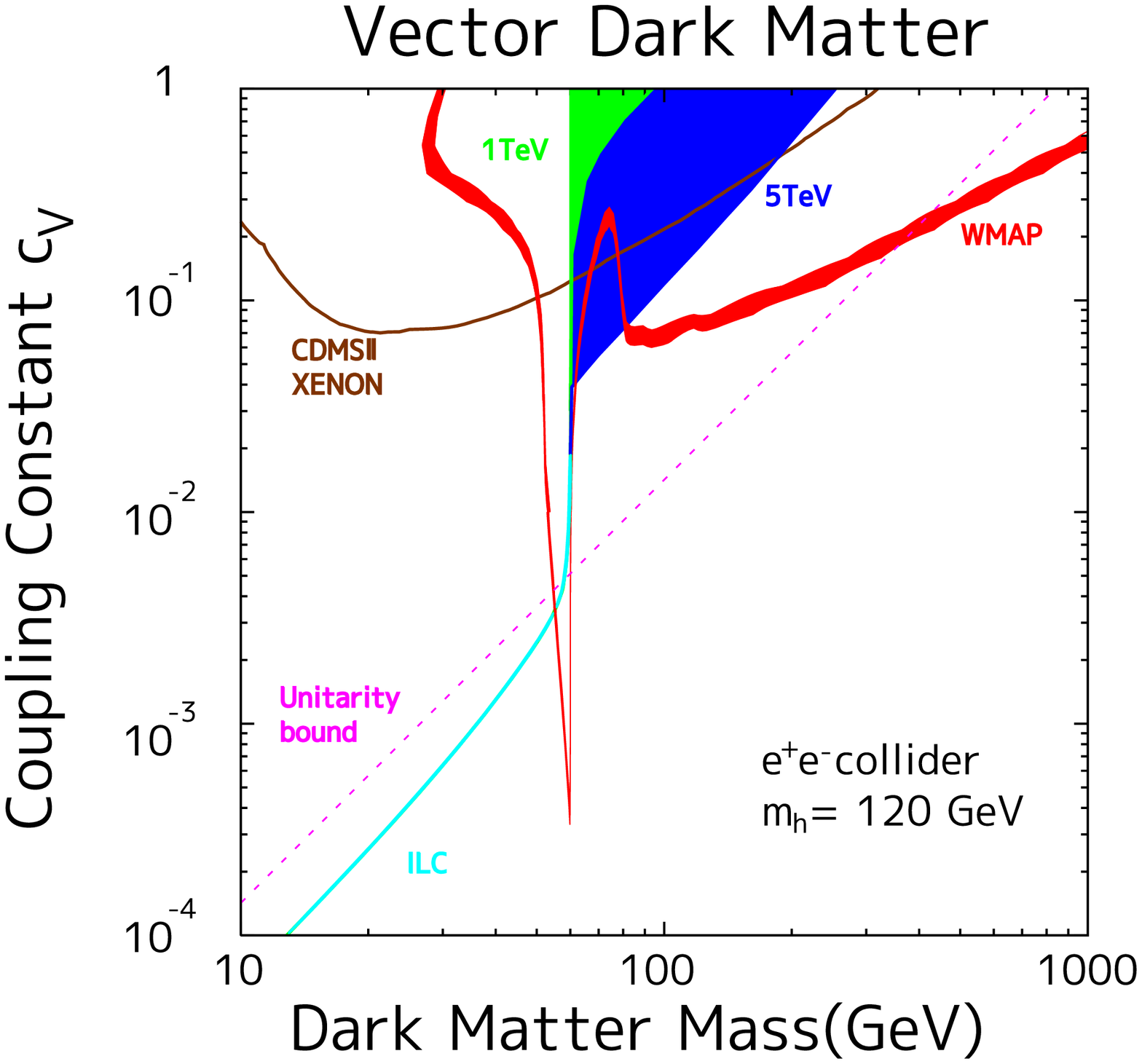}
 \end{center}
\caption{\small 
 Sensitivities to detect the dark matter signal at the ILC and CLIC. 
 The areas of $N_S/\sqrt{N_S+N_B} > 5$ at the $e^+e^-$ collider for $\sqrt{s} = 1$~TeV (green) 
 and 5~TeV (blue) with 1 ab$^{-1}$ data are shown with assuming $m_h = 120$~GeV.
 Constraints on direct detection experiments and the tree level unitarity 
 for dark matter are also shown.}
\label{fig:all}
\end{figure}
 
\hspace{12pt}In Fig.~\ref{fig:all}, we show the regions where the significance is larger than five in the plane of the coupling constant and the dark matter mass at $\sqrt{s} = 1$~TeV (green area) and 5~TeV (blue area). The mass of the Higgs boson is set to be 120~GeV and the integrated luminosity is assumed to be 1 ab$^{-1}$. For the region $m_D < m_h/2$, where the Higgs boson can decay into a pair of dark matters, the 3$\sigma$ line at $\sqrt{s} = 350$~GeV with the integrated luminosity 500~fb$^{-1}$ is shown by the cyan curve. In each figure, the allowed region which satisfies the WMAP data (3$\sigma$) is the red area. 

\hspace{12pt}First, in the figure on the left side of Fig.~\ref{fig:all}, the results for the scalar dark matter are shown. There is no overlap between the region of $N_S/\sqrt{N_S+N_B} > 5$ and that satisfying the WMAP data even at $\sqrt{s} = 5$~TeV. 
Second, in the figure on the center side of Fig.~\ref{fig:all}, the results for the fermion dark matter are shown. For the $e^+e^-$ collision at $\sqrt{s} = 1$~TeV, the area where $N_S/\sqrt{N_S+N_B} > 5$ and the WMAP data are both satisfied is very limited, while the area becomes wider at $\sqrt{s} = 5$~TeV. 
Finally, in the figure on the right side of Fig.~\ref{fig:all}, the results for the vector dark matter are shown. At $\sqrt{s} = 1$~TeV, $N_S/\sqrt{N_S+N_B} > 5$ and the WMAP data cannot be compatible, but a wide region of the overlap can be seen at the $\sqrt{s} = 5$~TeV. In particular, for the $m_h/2<m_D<100$~GeV, it can be seen that the vector dark matter with the coupling constant larger than $2$-$4\times 10^{-3}$ can be tested.

\section{Conclusions}

\hspace{12pt}We have investigated the possibility of detecting dark matter at TeV scale colliders 
in the Higgs portal dark matter scenario. 
We have considered three possible cases for spin of the dark matter (the scalar, fermion or vector dark matter) 
and we have studied pair production processes of the dark matter 
via weak boson fusion processes. 
If the dark matter particle is light, $m_D^{} < m_h^{}/2$, 
the Higgs boson decays into a pair of dark matter particles 
with a large branching ratio. 
Such an invisibly decaying Higgs boson can be explored 
at the LHC by the Higgs boson production process. 
We have found that a multi-TeV collider can be more useful to explore the dark matter 
in these models than the 1~TeV collider when the invisible decay of the Higgs boson 
into a pair of dark matters is kinematically forbidden. 
Suppose that the Higgs boson is found to be 120 GeV at the LHC 
and that in future an excess will be found for the signal of $e^+e^-$ 
plus missing energy above the background at the 5TeV linear collider. 
Our results tell us that such a signal would indicate the WIMP dark matter. 
The interaction is also determined to a considerable extent. 
Therefore, we conclude that 
by measuring this process at the multi-TeV linear collider, 
we may be able to extract the measure property of the WIMP dark matter 
such as its mass, spin, and coupling constants.  

\vspace{1.0cm}
\noindent
{\bf Acknowledgments}
\vspace{0.5cm}

\hspace{12pt}I would like to thank S.~Kanemura, S.~Matsumoto, N.~Okada and H.~Taniguchi 
for fruitful collaborations. 
This talk is partially supported by a JSPS grant in aid for specially promoted program 
"A global research and development program of a state-of-the-art detector system for ILC".

\end{document}